\documentclass[reqno]{amsart}
\usepackage{epsf}
\def\be{\begin{equation}}
\def\ee{\end{equation}}
\def\ba{\begin{array}}
\def\ea{\end{array}}
\def\bea{\begin{eqnarray}}
\def\eea{\end{eqnarray}}
\def\drm{{\mathrm d}}
\def\erm{{\mathrm e}}
\def\dps{\displaystyle}

\def\noi{\noindent}

\def\ov{\overline}

\begin{document}

\vspace{-4truecm} %
{}\hfill{DSF$-$9/2008}%

\vspace{1truecm}

\title{A theory of ferromagnetism by Ettore Majorana}

\author{S. Esposito}
\address{{\it S. Esposito}: Dipartimento di Scienze Fisiche,
Universit\`a di Napoli ``Federico II'' \& I.N.F.N. Sezione di
Napoli, Complesso Universitario di M. S. Angelo, Via Cinthia,
80126 Napoli ({\rm Salvatore.Esposito@na.infn.it})}%

\begin{abstract}
We present and analyze in detail an unknown theory of
ferromagnetism developed by Ettore Majorana as early as the
beginnings of 1930s, substantially different in the methods
employed from the well-known Heisenberg theory of 1928 (and from
later formulations by Bloch and others). Similarly to this,
however, it describes successfully the main features of
ferromagnetism, although the key equation for the spontaneous mean
magnetization and the expression for the Curie temperature are
different from those deduced in the Heisenberg theory (and in the
original phenomenological Weiss theory). The theory presented here
contains also a peculiar prediction for the number of nearest
neighbors required to realize ferromagnetism, which avoids the
corresponding arbitrary assumption made by Heisenberg on the basis
of known (at that time) experimental observations. Some
applications of the theory (linear chain, triangular chain, etc.)
are, as well, considered.
\end{abstract}

\maketitle



\section{Introduction}

\noi The microscopic interpretation of the magnetic properties of
bodies, related to the motion of electrons in atoms, appeared as
early as the beginning of the XX century, when P. Langevin
\cite{Langevin} provided a successful  theory of diamagnetism, by
evaluating the induced magnetization in presence of an applied
magnetic field in the framework of the Lorentz theory of the
electron. Subsequently the paramagnetism of materials was rightly
related to the fact that atoms or molecules can have permanent
dipole moments arising from the current loops generated by
orbiting electrons (proportional to the total angular momentum),
and thus acting as tiny magnetic shells. The correct
interpretation of the ferromagnetic phenomena, instead, did not
come out soon, and their understanding on an atomic basis was
achieved only after the advent of quantum mechanics.

The striking property to be explained was that the ferromagnetic
substances (like iron, cobalt, nickel, etc.) develop a spontaneous
magnetization only when cooled below a certain temperature $T_c$,
called the Curie point. Differently from an ordinary paramagnet,
above this critical value the temperature variation of the
magnetic susceptibility $\chi$ of a ferromagnet was found
empirically to follow the Curie-Weiss law, $\chi = C/(T-T_c)$
(where $C$ is a constant), whereas for a paramagnet the Curie law
$\chi=C/T$ holds.

\begin{figure}[t]
\begin{center}
\epsfysize=4truecm
\centerline{\epsffile{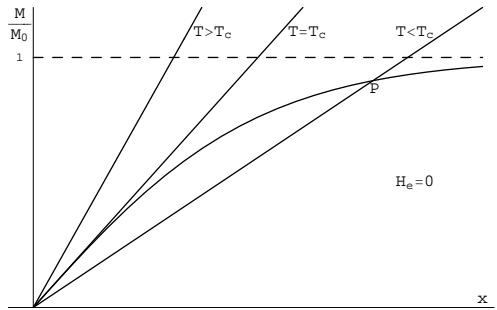}} %
\caption{Graphical solution of the implicit equation (\ref{1}),
obtained from the intersection of the two curves $M=M_0 \tanh x$
and $M=(kT/\mu \alpha) x$ for zero external magnetic field and
different values of the temperature. A non-vanishing spontaneous
magnetization (corresponding to the point P) exists only for
$T<T_c$.} %
\label{figweiss}
\end{center}
\end{figure}

Early works by J.A. Ewing, K. Honda and J. Okubo, and
others\footnote{For a description of these and other earlier
models see the review in Ref. \cite{Keehan}.} showed that many of
the peculiarities exhibited by ferromagnetic bodies could be
described by assuming a large potential energy between adjacent
microscopic (atomic or molecular) magnets. However, a first
important step towards the understanding of the phenomenon was
performed by P. Weiss in 1907 \cite{Weiss}, when he postulated
that the atomic magnets tend to be brought into parallel
orientation not only by an applied, external magnetic field $H_e$
but also by an inner field $H_i$, called by Weiss the ``molecular
field'', which is proportional to the magnetization $M$ of the
material: $H_i = \alpha M$. By introducing such an idea of an
inner field into the Langevin theory of the magnetism of classical
dipoles, Weiss was in fact able to account for most of the
phenomenology of ferromagnets \cite{Mermin}. If a dipole is
brought from outside and placed in the molecular field, it will be
aligned in the direction of this field (i.e. in the direction of
the magnetization itself), giving rise to a spontaneous ordering
of the elementary magnets. Such an ordering effect competes with
random thermal effects, with energy of the order of $kT$, trying
to flip the dipole away from  the ordered state, but Weiss showed
that (in the simple case that dipoles can align only in two
opposite direction), even in the absence of an external magnetic
field ($H_e=0$), a non-vanishing magnetization may arise,
satisfying the equation:
 \be
M = M_0 \tanh \frac{\mu \alpha M}{kT} \label{1}
 \ee
($\mu$ is the magnetic moment of the $N$ atoms per unit volume of
the body, and $M_0 = N \mu$).\footnote{Actually, in the original
formulation, the Langevin function $\cosh x - 1/x$ appeared in Eq.
(\ref{1}) instead of the function $\tanh x$, which was considered
only later by W. Heisenberg (see below). However, as it is easily
recognizable (as done by Heisenberg himself), the physical
interpretation of the phenomena does not change and no alteration
in the reasoning is required, so that we prefer to use the
mathematical form of Eq. (\ref{1}) for later convenience.} In
fact, apart from the trivial solution $M=0$, the implicit equation
(\ref{1}) admits also another non-zero solution for the
magnetization $M$, provided that $T < T_c = N \mu^2 \alpha / k$,
as it was seen graphically (see Fig. \ref{figweiss}). The magnetic
susceptibility was as well calculated, obtaining the
experimentally observed Curie-Weiss law (with $C=N \mu^2/k$), thus
finally showing the good agreement of the phenomenological Weiss
theory with the known properties of ferromagnetism.

However, despite such a success, the question about the origin of
the local field remained unanswered, and Weiss himself realized
that it could not arise from the magnetic interactions of the
magnetic moments of the atoms or molecules, these being extremely
small. In fact, the value of the proportionality coefficient
$\alpha$ should be assumed as high as of the order of $10^4$,
against the value of $4 \pi/3$ calculated under ordinary
electromagnetic assumption \cite{VanVleck}. On the other hand,
classical electrostatic forces may lead to interactions of the
right order of magnitude, but do not give a linear proportionality
between the Weiss molecular field $H_i$ and the magnetization $M$
\cite{Debye}.

The solution of this puzzle had to wait until the end of 1920s,
when W. Heisenberg \cite{Hei28} showed that the forces leading to
ferromagnetism are purely quantum mechanical in nature, being due
to electron exchange. The key idea was to show how strong
non-magnetic forces between electrons that favor spin alignment
(and, thus, the alignment of the elementary magnets) arise from
the quantum mechanical exchange interactions, in the same way as
they produce level splittings between electrons in singlet (para-)
and triplet (ortho-) states of two-electron atoms: ``the empirical
phenomenon that ferromagentism presents is very similar to the
situation we met earlier in the case of the helium atom.''
\cite{Hei28}. Heisenberg, indeed, began in 1926 \cite{Hei26} to
explore how the so-called ``exchange term'' appears in a system of
atoms with many electrons satisfying the Pauli exclusion
principle. Exchange forces are electrostatic in nature but,
because of the constraints imposed by the Pauli principle, they
are formally equivalent to a tremendously large coupling between
spins, as seemingly required by the Weiss phenomenological theory.
With a suitable assumption (a {\it positive} exchange integral),
Heisenberg showed that the state in which spins are aligned in the
same direction is energetically favorable, helping in the ordering
of the spins in a certain direction, thus proving his model to be
well fitted to describe ferromagnetism.

This result was, actually, achieved by following the statistical
model built up by E. Ising few years before \cite{Ising}. In the
Ising model (or, rather, the Lenz-Ising model \cite{Brush}) of a
ferromagnetic body, $N$ elementary magnets of moment $\mu$ are
assumed to be arranged into a regular lattice, each of them having
only two possible orientations, as in the Weiss theory. The
magnets experience only short-range interactions on each other,
the interaction energy (assumed to be a constant) being
non-vanishing and negative (positive) only for each pair of
neighboring magnets of same (opposite) direction. This interaction
tends, then, to make neighboring spins the same (an additional
energy $\pm \mu H_e$ is present for an applied magnetic field
$H_e$), so that in principle this model could be suitable for
describing ferromagnetic phenomena. The problem is just to find an
analytic expression for the statistical mechanical partition
function
 \be
Z = \sum_{\rm config.} \erm^{- \frac{E}{kT}}, \label{2}
 \ee
where $E$ is the total energy of the system (and the sum is over
any configuration of the system), from which all the thermodynamic
quantities of interest (including the mean magnetization) can be
derived. This was effectively done by Ising for magnets arranged
into a linear chain, but he ``succeeded in showing that the
assumption of directed, sufficiently great forces between two
neighboring atoms of a chain is not sufficient to explain
ferromagnetism'' \cite{Hei28}, since no non-zero magnetization is
predicted at any temperature. This occurs since the ordered state
is unstable against random thermal fluctuations tending to destroy
spin alignment, and Ising showed (incorrectly) that such negative
results applies also in three dimensions.

This problem was not properly solved by Heisenberg in 1928,
although his different (non constant) form of the interaction
energy depends not only on the arrangement of the elementary
magnets, but also on the speed with which they exchange their
places. It was correctly addressed only in 1936 by R. Peierls
\cite{Peierls} who, against quite a commonly expressed opinion
``that the solution of the three-dimensional problem could be
reduced to that of the linear model and would led to similar
results'', finally showed that the above-mentioned statistical
model of ferromagnetism gives non-zero spontaneous magnetization
at finite temperature in two (or more) dimensions.

Returning to the Heisenberg theory, his derivation of the Weiss
formula (\ref{1}) proceeded through standard calculations on the
partition function (\ref{2}). The first step was to find the
distribution of the energy levels of $2n$ localized electrons (one
valence electron per atom) to first order in the coupling by
evaluating the exchange integral between each electron and its
nearest neighbors. Given the exponential decrease of the exchange
integrals, the couplings among more distant electrons are, in
fact, negligible. Formal group-theoretic methods were used to
evaluate the energy levels and, in order to calculate the
partition function, their fluctuations were neglected. This last
assumption effectively meant the replacement of the energy levels
of the spin multiplets of the microscopic magnets forming a given
total spin $\tt S$ by the average energy corresponding to $\tt S$.
Subsequently, with a ``somewhat arbitrary'' assumption, Heisenberg
introduced a Gaussian distribution of energies for $\tt S$ around
its mean energy $E_0$ with variance $\ov{\Delta E_0^2}$; as he
himself wrote, ``it is of course only to be expected that the
temporary theory sketched here offers but a qualitative scheme
into which ferromagnetic phenomena will perhaps be later
incorporated'' \cite{Hei28}.

Subsequent improvements, both on the mathematical formalism and on
the comprehension of several physical effects, came out between
the end  of 1920s and the beginning of 1930s, mainly due to F.
Bloch, J. Slater and H.A. Bethe. Firstly, Bloch and Slater
concentrated (independently) on the understanding of the physical
basis of the Heisenberg model. While considering a free electron
gas (instead of the bound electrons contemplated by Heisenberg in
his Heitler-London-like \cite{HeitlerLondon} approach) for
studying the role of conduction electrons in ferromagnets (for
obtaining its thermal properties), Bloch avoided Heisenberg's
Gaussian assumption and found a contribution coming out ``when two
electrons are close together, in the neighborhood of the same
atom'' \cite{Bloch1929}. Such an important effect (the exchange
integral can become negative, this decreasing the possibility to
have ferromagnetism), in fact, cannot be taken into account in the
Heitler-London-like approach used by Heisenberg. Furthermore Bloch
replaced the difficult group-theoretic method employed by
Heisenberg with the simpler determinantal method introduced
shortly before by Slater in his theory of complex spectra
\cite{Slater1929}. The great advantage of writing a many-electron
wavefunction in the form of a Slater determinant was just to take
into account from the very beginning the correct antisymmetry
properties imposed by the Pauli principle, thus making sure that
no two electrons in the system can be found in identical quantum
states. This novel approach directly led Bloch to discover the
so-called ``spin waves'' \cite{Bloch1930}, i.e. the states of the
system corresponding to single or few spin flips in the fully
aligned ground state, in the region of low temperatures, where
both the Weiss phenomenological theory and the Heisenberg
calculations couldn't be applied.

All these results by Heisenberg, Bloch and Slater were reviewed
(and criticized) by W. Pauli in his comprehensive review of
magnetism delivered at the October 1930 Solvay conference in Paris
\cite{PauliSolvay}.

This meeting was also attended by E. Fermi who, very likely, was
responsible of the spreading of the novel results about
ferromagnetism within his group in Rome. Although Fermi and his
close collaborators never worked and published on this argument,
several people visiting the Institute of Physics in Rome in the
early 1930s did, including, primarily, H.A. Bethe
\cite{Bethe1931}. In Rome, he calculated eigenvalues and
eigenfunctions (at first approximation) for any value of the
resulting magnetic moment of the whole crystal (in the
one-dimensional case of a linear chain), by introducing the famous
{\it Bethe ansatz} for the case of two interacting spin waves.
Other minor (to some extent) contributions came out from D.R.
Inglis \cite{Inglis1}-\cite{Inglis3}, who visited Rome in 1932,
and from a member of the extended Fermi group, G. Gentile, who
provided \cite{Gentile} a thorough justification of Bloch's
results on elementary groups ({\it Uebergangsgebiet}) of aligned
spins inside the ferromagnet \cite{Bloch1932}. Inglis, instead,
considered the problems about the applicability of Heisenberg's
approximation on the Gaussian distribution of energy eigenvalues
\cite{Inglis1} and, later, the effect of the spin-orbit
interactions in ferromagnetic anisotropy \cite{Inglis2} and the
problem of the lack of orthogonality for the electronic
wavefunctions of two neighboring atoms \cite{Inglis3}.

Here we are not interested to the quite irrelevant question about
the possible origin of such works within the Fermi group in
Rome,\footnote{Just to quote a different example, we point out
that Bloch and Gentile worked together in Leipzig when, in 1931,
they were both at the Institute headed by Heisenberg} but rather
report on an unknown but very interesting contribution on
ferromagnetism by another distinguished member of that group, E.
Majorana, who much contributed to the achievements of several
results in different fields of research, results that were often
related to other people (for a recent review, see \cite{ADP} and
references therein). As quite usual for Majorana, many of these
achievements were never published by him (and, in most cases, even
his friends and colleagues were not aware of them), so that they
are practically unknown. Fortunately enough, however, Majorana's
calculations have been preserved in his personal notebooks
\cite{Volumetti} \cite{Quaderni} and we are, thus, able to
reconstruct, in particular, Majorana's contribution on
ferromagnetism \cite{Quaderni}. This is presented in detail in the
following section, while we comment on its relevance and some
intriguing aspects in Section 3. Finally, our conclusions and main
results are summarized in Section 4.

\section{Majorana theory of ferromagnetism}

\noi The Majorana contribution begins with focusing on a set of
$n$ atoms, considered as magnetic dipoles (or, briefly, {\it
spins}), arranged into a given geometric array at locations $q_1,
q_2, \dots q_n$. Among these $n$ elements, he assumes that a
number $i$ of them has a spin parallel to a given direction ({\it
spin-up}), while the remaining $n-i$ elements are antiparallel to
that direction ({\it spin-down}). Spin-up elements are
characterized by the quantum numbers:
\[
r_1 , r_2, \ldots r_i \qquad \uparrow \, \uparrow \, \ldots \,
\uparrow,
\]
while those with spin-down are characterized by:
\[
r_{i+1}, r_{i+2}, \ldots r_n \qquad \downarrow \, \downarrow
\ldots \, \downarrow .
\]
The wavefunctions $A (r_1 \ldots r_i | r_{i+1} \ldots r_n)$
describing such a system have the form of a Slater determinant; in
the following expression, the $\delta$-functions (or, more
appropriately, the Kronecker symbols) $\delta(s_k \mp 1)$
guarantee that spin-up/spin-down elements have $s=\pm 1$ spin (in
some unit), respectively:
 \be %
A (r_1 \ldots r_i | r_{i+1} \ldots r_n) = \left|
\begin{array}[c]{lll}
\psi_{r_1} (q_1) \delta (s_1 -1) & \ldots & \psi_{r_1}
(q_n) \delta (s_n -1) \\ \ldots & & \\
\psi_{r_i} (q_1) \delta (s_1 -1) & \ldots & \psi_{r_i} (q_n)
\delta (s_n -1) \\ \\ \psi_{r_{i+1}} (q_1) \delta (s_1 +1) &
\ldots & \psi_{r_{i+1}} (q_n) \delta (s_n +1) \\ \ldots & & \\
\psi_{r_n} (q_1) \delta (s_1 +1) & \ldots & \psi_{r_{n}} (q_n)
\delta (s_n +1) \end{array} \right| . \label{eq1}
 \ee %
Since there are
\[ \tau = \left( \ba{c} n \\ i \ea \right)
= \frac{n!}{i! (n-i)!} \] ways to have $i$ spin-up and $n-i$
spin-down on a total of $n$ elements (the order of $r_1 \ldots
r_i$ or $r_{i+1} \ldots r_n$ is not important), the wavefunctions
describing the system are the following:
 \bea
& & A_i (r^1_1, r^1_2 \ldots r^1_1|r^1_{i+1}, r^1_{i+2} \ldots
r^1_n ) , \nonumber \\
& & \ldots \label{eq2} \\
& & A_\tau (r^\tau_1, \ldots \ldots r^\tau_1|r^\tau_{i+1},
r^\tau_n) . \nonumber
 \eea
Denoting with $H$ the hamiltonian operator describing the
interaction acting on each element, the electrostatic interaction
potential $V_0$ is given by:
 \be
V_0 = \int H \psi_1 (q_1) \ov{\psi}_1 (q_1) \psi_2 (q_2)
\ov{\psi}_2 (q_2) \ldots \psi_n (q_n) \ov{\psi}_n (q_n) \ \drm q_1
\ldots \drm q_n . \label{eq3}
 \ee
The exchange energy between states characterized by $r$ and $s$
quantum numbers (or, according to Majorana's language, the
exchange energy between $r$ and $s$ orbits) is instead:
 \be
V_{rs} = \int \frac{e^2}{|q_1 - q_2|} \psi_r (q_1) \ov{\psi}_s
(q_1) \ov{\psi}_r (q_2) \psi_s (q_2) \ \drm q_1 \drm q_2 ,
\label{eq4}
 \ee
with $V_{rs} = V_{sr}$. The energy eigenvalues are, then:
 \be
H_{mm} = V_0 - \sum_{r<s} V_{rs} + \sum_{r=r_1^m}^{r_i^m} \
\sum_{s=r_{i+1}^m}^{r_n^m} V_{rs} \label{eq5}
 \ee
(the second term describing the action of given elements on
themselves, while the third one that of given elements on other
elements), while for $m \neq n$:
 \be
H_{mn} = \left\{ \begin{array}{lcl} - V_{rs}, & &
\mbox{for a transition from $A_m$ to $A_n$ by exchanging} \\
& & \mbox{the opposite intrinsic orientation in the orbits} \\
& & \mbox{$\psi_r$ and $\psi_s$,}
\\ & & \\
0, & & \mbox{for the other cases.}
\end{array} \right.
\label{eq6}
 \ee
Majorana now assumes that ($\epsilon$ is a constant)
 \be
V_{rs} = \left\{ \begin{array}{lll}
\varepsilon, & & \mbox{ for neighbor atoms}, \\
& & \\
0, & & \mbox{ for distant atoms},
\end{array} \right.
 \label{eq7}
 \ee
each atom having a number $a$ of neighbor atom. The energy change
in the ferromagnetic case is:
 \bea
E &=& H - V_0 + \sum_{r<s} V_{rs} \label{eq8} \\
&=& H - V_0 + \frac{n a}{2} \varepsilon \nonumber
 \eea
whose eigenvalues are:
 \bea
E_{mm} &=& \sum_{r_1^m}^{r_i^m} \sum_{r_{i+1}^m}^{r_n^m} V_{rs}
\label{eq9} \\
&=& N_m \ \varepsilon \nonumber
 \eea
and, for $m \neq n$:
 \be
E_{mn} = \left\{ \begin{array}{l} - V_{rs}, \\ \\ 0. \end{array}
\right. \label{eq10}
 \ee
Let $y=y(N)$ be the number of configurations in which $N$ spins
can be arranged (at a given temperature) among each other, subject
to the normalization condition:
\[
\sum_N y(N) = \left( \ba{c} n \\ i \ea \right) .
\]
In order to evaluate the partition function of the system, the
precise determination of the energy levels (\ref{eq8}),
corresponding to a given spin state of the system, is required,
but this is virtually impossible due to the large number of
elements involved. In his theory, Heisenberg therefore made the
approximation that the discrete succession of such energy levels
can be replaced by a continuum, and that such a continuum is
distributed according to the Gaussian error law around the average
energy for the given spin state of the system. In his notebooks,
Majorana makes a similar assumption: ``Can we consider $E$ as
diagonal, in a statistical sense? Let us assume that it can be.''
The distribution function, now subject to the normalization
condition
 \be
\int_{-\infty}^{+\infty} \! y \ \drm N = \left( \ba{c} n \\ i \ea
\right) ,
 \label{eq11}
 \ee
is then assumed to be picked around a pronounced maximum
 \be
y_0 = y(N_0)
 \label{eq12}
 \ee
as in the case of a Gaussian function. This means that, with a
maximum probability, we have $y_0$ quantities $A$ in (\ref{eq2})
corresponding to $N_0$.

\begin{figure}[t]
\begin{center}
\epsfysize=4truecm
\centerline{\epsffile{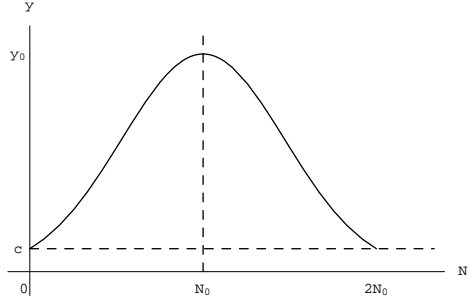}} %
\caption{A sketch of the continuous distribution function $y(N)$.} %
\label{fig1}
\end{center}
\end{figure}

Then, Majorana continues: ``In each of the quantities $A$ we
exchange randomly'' (due to temperature effects) ``an orbit
$\uparrow$ with an orbit $\downarrow$; the quantities $A$ change
into quantities $B$:
 \be
\begin{array}{lcl}
A_1 & \longrightarrow & B_1, \\
\ldots & & \\
A_\tau & \longrightarrow & B_\tau .
\end{array}
 \label{eq13}
 \ee
Statistically, the set of quantities $B$ coincides with that of
the quantities $A$. If we perform the transformation (\ref{eq13}),
the quantities $B$ corresponding to the $y_0$ quantities $A$ will
be distributed between $N_0-2a$ and $N_0+2a$.''

Let $p_{2r}$ be the probabilities that one among the mentioned $B$
quantities corresponds to $N=N_0+2r$, with $-a \leq r \leq a$. The
``average increment'' in the number of aligned spin is
 \be
\ov{\Delta N_0} = \sum_{r=-a}^a 2r \ p_{2r}
 \label{eq14}
 \ee
and can be evaluated as follows. On average, one of the $i$
spin-up elements has $N_0/i$ spin-down and $a - N_0/i$ spin-up
elements as neighbors, while one of the $n-i$ spin-down elements
has $N_0/(n-i)$ spin-up and $a - N_0/(n-i)$ spin-down elements as
neighbors. By performing the mentioned transformation
(\ref{eq13}), we have
 \be
\ov{\Delta N_0} = 2a - 2 \left( \frac{N_0}{i} + \frac{N_0}{n-i}
\right) .
 \label{eq15}
 \ee
Since we have a total of $\ell = 2a$ spins around $N_0$, the
probability that a given element be a spin-up is assumed by
Majorana to be
 \be
q = \frac{1}{2} + \frac{\ov{\Delta N_0}}{4a} ,
 \label{eq16}
 \ee
while for the probability to have a spin-down:
 \be
1 - q = \frac{1}{2} - \frac{\ov{\Delta N_0}}{4a} .
 \label{eq17}
 \ee
Thus, the probability to find $m=a+r$ spin-up elements that
alternate every $\ell -m = a-r$ spin-down elements is given by
 \be
\left( \ba{c} \ell \\ m \ea \right) q^m (1-q)^{\ell -m},
 \label{eq18}
 \ee
since $q$ is evidently a random binomial variable. By replacing in
(\ref{eq18}) the corresponding quantities in (\ref{eq16}),
(\ref{eq17}), we then get:
 \be
p_{2r} = \left( \frac{1}{2} + \frac{\ov{\Delta N_0}}{4a}
\right)^{a+r} \left( \frac{1}{2} - \frac{\ov{\Delta
N_0}}{4a}\right)^{a-r} \frac{(2a)!}{(a-r)! \, (a+r)!}
 \label{eq19}
 \ee
(such a law for the probability is called ``normal'' by Majorana).

According to statistical mechanics, the ratio of the probabilities
(\ref{eq16}) and (\ref{eq17}) is just given by the Boltzmann
factor:
 \be
\frac{q}{1-q} = \erm^{-\frac{E}{kT}} .
 \label{eq20}
 \ee
However, Majorana proceeds in a slightly different way. ``Assuming
that, for a restricted range,
 \bea
 & & y (N_0 +1) = y_0 \ \erm^K, \nonumber \\
 & & \ldots  \label{eq21} \\
 & & y(N_0 \pm a) = y_0 \ \erm^{\pm K a}, \nonumber
 \eea
the condition that $y(N)$ does not change while we pass from
quantities $A$ to quantities $B$ can be expressed as:
 \be
 \sum_{r=-a}^a \,\,\, p_{2r} \ \erm^{-2Kr} =1 ,
 \label{eq22}
 \ee
that is solved by:
 \be
 \erm^K = \frac{\dps 1 + \frac{\Delta N_0}{2a}}{\dps 1-
 \frac{\Delta N_0}{2a}}, \quad \mbox{''}
 \label{eq23}
 \ee
which is just Eq. (\ref{eq20}), $\erm^K$ being the Boltzmann
factor. In the continuum limit, $K$ may be replaced by
$y^\prime/y$. ``It follows that, by considering $y$ as a
continuous function of $N$:
 \be
\frac{y'}{y} = \log \frac{\dps 2 - \frac{N}{a} \left( \frac{1}{i}
+ \frac{1}{n-i}\right) }{\dps \frac{N}{a} \left( \frac{1}{i} +
\frac{1}{n-i} \right)} ,
 \label{eq24}
 \ee
and setting
 \be
\alpha = \frac{1}{a} \left( \frac{1}{i} + \frac{1}{n-i} \right) =
\frac{n}{a \ i \ (n-i)} ,
 \label{eq25}
 \ee
we have:
 \be
\frac{y'}{y} = \log \left( \frac{2}{\alpha N} - 1 \right) . \quad
\mbox{''}
 \label{eq26}
 \ee
Such a differential equation is solved by means of separation of
variables, with the change of variable
 \be
t = \frac{2}{\alpha N} -1 .
 \label{eq27}
 \ee
After few calculations, Majorana finds:
 \be
y = c \left(\frac{2}{\alpha N} \right)^{\frac{2}{\alpha}} \left(
\frac{2 }{\alpha N} - 1 \right)^{-\frac{2}{\alpha} +N}
 \label{eq28}
 \ee
or
 \be
y = c \left( \frac{2}{2 - \alpha N} \right)^{\frac{2}{\alpha}}
\left( \frac{2}{\alpha N} - 1 \right)^N .
 \label{eq29}
 \ee
The constant $c$ corresponds to:
 \be
y(0) = y \left( \frac{2}{\alpha} \right) = c
 \label{eq30}
 \ee
(a limit procedure is, obviously, adopted), and may be calculated
from the normalization condition (\ref{eq11}). Here, in order to
evaluate the integral, Majorana approximates the function $y(N)$
to a Gaussian curve,
 \be
y \simeq  y \left( \frac{1}{\alpha} \right) \erm^{- \alpha \left(
N - \frac{1}{\alpha} \right)^2}
 \label{eq31}
 \ee
(note that $y(N)$ takes its maximum at $N=1/\alpha$, and
$y(1/\alpha) = 2^{2/\alpha}c$), obtaining ($N'=N-1/\alpha$):
 \[
\int_{-\infty}^{+\infty} y \ \drm N \cong \int_{-\infty}^{+\infty}
y \left( \frac{1}{\alpha} \right) \erm^{-\alpha N'^2} \drm N' =
\sqrt{\frac{\pi}{\alpha}} \ y \left( \frac{1}{\alpha}\right) =
\sqrt{\frac{\pi}{\alpha}} \ c \ 2^{\frac{2}{\alpha}}.
 \]
From this, we get:
 \be
c = \left( \ba{c} n \\ i \ea \right) \left( \frac{1}{2}
\right)^{\frac{2}{\alpha}} \sqrt{\frac{\alpha}{\pi}} .
 \label{eq32}
 \ee
The solution found is, then, worked out with a specific numerical
example, for $n=10$, $i=3$, $a=4$ (corresponding to $\alpha =
5/42$, $2/\alpha = 16.8$ and $c = 0.0002046$), and a numerical
tabulation for it is reported in the notebook. Majorana also
pushes further the approximation (\ref{eq31}) by expanding, in a
Taylor series, $\log y$ rather than directly $y$, finally
obtaining:
 \be
\log y  =  \ov{k} - \alpha N'^2 - \frac{1}{6} \alpha^3 N'^4 -
\frac{1}{15} \alpha^5 N'^6 - \frac{1}{28}\alpha^7 N'^8 -
\frac{1}{45} \alpha^9 N'^{10} - \ldots
 \label{eq33}
 \ee
with $\ov{k}$ a given constant.

At a given temperature $T$, the partition function of the system
is evaluated from the statistical term
\[ y \ \erm^{- \frac{E}{kT}}, \]
which now replaces the distribution function $y$. The Boltzmann
factor may be written as
 \be
\erm^{- \frac{E}{kT}} = \erm^{- \frac{N \epsilon}{kT}} = \erm^{- L
\, N}
 \label{eq34}
 \ee
with
 \be
L = \frac{\epsilon}{kT}
 \label{eq35}
 \ee
The maximum of the probability distribution function
(corresponding to $N=N_0$) may be evaluated by taking
\[
\frac{\drm ~}{\drm N} \log \left(  y \ \erm{- \frac{E}{kT}}
\right) = 0
\]
(and checking, as done by Majorana, the sign of the second
derivative). From (as above, $N'=N-1/\alpha$):
 \bea
\log \left( y \ \erm^{- \frac{W}{kT}} \right) = \ov{k} -
\frac{1}{\alpha} - L N' - \frac{1}{\alpha} \log (1 - \alpha^2
N'^2) + N' \log \frac{1 - \alpha N'}{1 + \alpha N'} , \label{eq36}
 \eea
 \bea
\frac{\drm ~}{\drm N} \log \left( y \ \erm^{-\frac{W}{kT}} \right)
= -L + \log \left( \frac{2}{\alpha N} - 1\right), \nonumber
 \eea
we obtain:
\[
\frac{2}{\alpha N_0} - 1 = \erm^L
\]
or
 \be
N_0 = \frac{2}{\alpha (\erm^L +1)}
 \label{eq37}
 \ee
(note that for $L=0$ one re-obtains $N_0=1/\alpha$). By setting
$N=N_0$ in (\ref{eq29}) we have
 \be
y_0 = c \left( \frac{\erm^L +1}{\erm^L} \right)^{\frac{2}{\alpha}}
\erm^{LN_0}
 \label{eq38}
 \ee
and
 \be
y_0 \ \erm^{- \frac{E_0}{kT}} =  c \ \left( \frac{\erm^L
+1}{\erm^L} \right)^{\frac{2}{\alpha}} \erm^{LN_0} \ \erm^{- LN_0}
= c \left( \frac{\erm^L +1}{\erm^L} \right)^{\frac{2}{\alpha}} .
 \label{eq39}
 \ee
More in general, by expanding Eq. (\ref{eq36}) in a Taylor series
up to second order terms in $(N-N_0)$ and using (\ref{eq37})
(\ref{eq39}) we get:
 \be
y \ \erm^{-\frac{E}{kT}} = c \left( \frac{\erm^L +1}{\erm^L}
\right)^{\frac{2}{\alpha}} \erm^{- \frac{(\erm^L+1)^2}{4\erm^L}
\alpha (N-N_0)^2} .
 \label{eq40}
 \ee
From the expression (\ref{eq32}) we finally obtain:
 \bea
\int y \ \erm^{- \frac{E}{kT}} \drm N & = & \left(
\begin{array}{c} n \\ i \end{array} \right) \left( \frac{1 +
\erm^{-L}}{2} \right)^{\frac{2}{\alpha}}
\frac{2\sqrt{\erm^L}}{\erm^L +1} \nonumber \\
& = & \left( \begin{array}{c} n \\ i \end{array} \right) \left(
\frac{1 + \erm^{-L}}{2} \right)^{a\frac{2i (n-i)}{n}}
\frac{2\sqrt{\erm^L}}{\erm^L +1}
 \label{eq41}
 \eea
(in the last passage, Eq. (\ref{eq25}) has been used).

At this point, Majorana considers the case where a magnetic field
$\tt H$ is applied to the system, and puts
 \be
M = \frac{\mu {\tt H}}{k T}
 \label{eq42}
 \ee
($\mu$ denoting the magnetic moment of a given atom). The
additional magnetic energy is ruled by the net number of spins
aligned along the direction of $\tt H$ ($n_{\uparrow}$) or
counter-aligned along the same direction ($n_{\downarrow}$):
 \be
- \left( n_{\uparrow} - n_{\downarrow} \right) \mu {\tt H} = -
(2i-n) \mu {\tt H} .
 \label{eq43}
 \ee
Note that the local magnetization, in units such that the absolute
saturation value is one, is given by
 \be
S = \frac{n_{\uparrow} - n_{\downarrow}}{n_{\uparrow} +
n_{\downarrow}} = \frac{2i-n}{n} ,
 \label{eq44}
 \ee
so that the mean magnetization of the system, i.e. ``the ratio
${\sf S}$ between the magnetic moment under the influence of the
field $\tt H$ and the saturation magnetic moment'', is:
 \bea
{\sf S} &=& \frac{\dps \sum_i \frac{2i-n}{n} \int y \ \erm^{-
\frac{E}{kT}} \erm^{M(2i-n)} \drm N}{\dps \sum_i \int y \ \erm^{-
\frac{E}{kT}} \erm^{M(2i-n)} \drm N} \nonumber \\
&=& \frac{\dps \sum_i \frac{2i}{n} \left( \begin{array}{c} n \\ i
\end{array} \right) \left( \frac{1 + \erm^{-L}}{2} \right)^{a
\frac{2i (n-i)}{n}} \erm^{M (2i-n)}}{\dps \sum_i \left(
\begin{array}{c} n \\ i \end{array} \right) \left( \frac{1 +
\erm^{-L}}{2}\right)^{a \frac{2i (n-i)}{n}} \erm^{M (2i-n)}} -1 .
\label{eq45}
 \eea
The (logarithm of the) partition function
 \be
Z = \sum_{n_\uparrow, n_\downarrow} p({\sf S}) ,
 \label{eq46}
 \ee
where $p({\sf S})$ is the probability to have the magnetization
${\sf S}$, in the continuum limit may then be evaluated from
 \bea
\log \int \erm^{- \frac{E}{kT}} \ \erm^{M (2 i - n)} y \ \drm N
&=& a \frac{2i (n-i)}{n} \log \frac{1+\erm^{-L}}{2} + 2Mi - i \log
i \nonumber \\
& & - (n-i) \log (n-i) + \mbox{constant terms},
 \label{eq47}
 \eea
where the Stirling approximation for the factorials (in the
binomial coefficients)
\[ \log n! \simeq n \log n - n \]
has been used. Here the omitted terms do not depend on the number
$i$ of spin-up elements. The condition for having a maximum in the
magnetization (in the case of ferromagnetism) may then be obtained
``by taking the derivative of (\ref{eq47}) with respect to $i$ and
equating the result to $0$'':
\[ a \frac{2 (n-2i)}{n} \log \frac{1- \erm^{-L}}{2} + 2M - \log i
+ \log (n-i)  =0 , \]
\[ \log \frac{i}{n-i} = 2M + a \frac{2(n-2i)}{n} \log \frac{1 +
\erm^{-L}}{2} , \] %
or, by using (\ref{eq44}),
 \be
 \log \frac{1+S}{1-S} =2 \frac{\mu {\tt H}}{kT} + 2a S \log \frac{2}{1
+ \erm^{- \frac{\epsilon}{kT}}} .
 \label{eq48}
 \ee
For small $\tt H$ and large $T$, this reduces to:
 \be
 2 S = 2 \frac{\mu {\tt H}}{kT} + 2a S \log \frac{2}{1
+ \erm^{- \frac{\epsilon}{kT}}} .
 \label{eq49}
 \ee
``For $T$ lower than the Curie point, for a given value of $\tt H$
there exist 2 values of $S$ which, for not extremely high $\tt H$,
are practically equal and opposite.''

Let now $s>0$ be the local magnetization of the system for ${\tt
H} = 0$:
 \be
 \log \frac{1+s}{1-s} = 2a s \log \frac{2}{1 + \erm^{- \frac{\epsilon}{kT}}} .
 \label{eq50}
 \ee
The influence of the magnetic field $\tt H$ on the magnetization
$S$ may be evaluated by putting $S=s+ \Delta s$ in Eq.
(\ref{eq48}); at first order in $\Delta s$ we have:
 \be
\Delta s = \frac{\dps \frac{\mu {\tt H}}{kT}}{\dps \frac{1}{1-
s^2} - a \log \frac{2}{1-\erm^{-\frac{\epsilon}{kT}}}}.
 \label{eq51}
 \ee

\

\section{Some implications and further contributions}

\noi The main body of the Majorana contribution about
ferromagnetism, present in his {\it Quaderni}, is all that
reported in the previous section (apart from some omissions of
scratch calculations and few other interesting points; see below).
Here we point out some peculiar features of the Majorana theory,
directly obtainable from the calculations reported but not
explicitly mentioned in the notebooks. Nevertheless, even though
in a form probably different from that given below, it is likely
that Majorana was aware of them.

Let us start with stressing that, besides the trivial solution
$S=0$, Eq. (\ref{eq48}) admits also another non-vanishing solution
for ${\tt H} = 0$, depending on the temperature. In fact, the
equation:
 \be \label{eq53}
\log \frac{1+S}{1-S} = \eta \, S, \qquad \qquad \eta = 2a \log
\frac{2}{1 + \erm^{- \frac{\epsilon}{kT}}}
 \ee
has a $S\neq 0$ solution if the following condition is fulfilled:
 \be \label{eq55}
2a \log \frac{2}{1 + \erm^{- \frac{\epsilon}{kT}}} > 2, \qquad
\mbox{or} \qquad \erm^{- \frac{\epsilon}{kT}}< 2 \erm^{-
\frac{1}{a}} -1 .
 \ee
This condition is non trivially satisfied only if $2 \erm^{-
\frac{1}{a}} -1 > 0$, that is
 \be \label{eq56}
a > \frac{1}{\log 2} \simeq 1.44.
 \ee
In such a case, the condition (\ref{eq55}) gives:
 \be \label{eq57}
T < \frac{\epsilon}{\displaystyle k \left| \log \left( 2 \erm^{-
\frac{1}{a}} -1 \right) \right| } \equiv T_c .
 \ee
Thus, if the temperature is lower than the critical value given in
Eq. (\ref{eq57}), the model studied by Majorana predicts a
spontaneous (${\tt H} = 0$) magnetization $S \neq 0$, i.e.
ferromagnetism, exactly as in the phenomenological Weiss theory,
although in a form different from that coming out from Eq.
(\ref{1}). In addition to this, and as a necessary condition for
it to occur, the inequality (\ref{eq56}) must be fulfilled, that
is the integer number of nearest neighbor atoms has to be at least
2, a prediction not contained in the Heisenberg model (see the
Introduction). Note also that, in the Majorana view, the quantity
$a$ effectively may count the number of physical dimensions (see
below) of the crystal structure of the ferromagnetic body and, in
this respect, such a result anticipates by several years what
definitively proved later by Peierls \cite{Peierls}.

\begin{figure}[t]
\begin{center}
\epsfysize=4truecm
\centerline{\epsffile{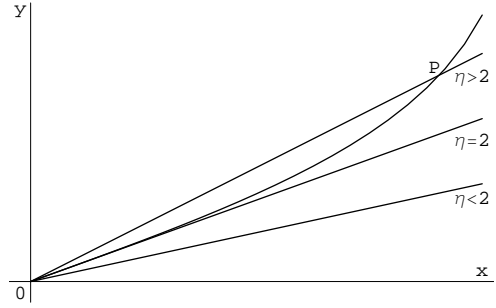}} %
\caption{Graphical solution of the implicit equation (\ref{eq53}),
obtained from the intersection of the two curves $y = \log
\frac{1+x}{1-x}$ and $y = \eta x$ for different values of the
parameter $\eta$. A non-vanishing spontaneous magnetization
(corresponding to the
point P) exists only for $\eta >2$, that is $a \geq 2$ and $T<T_c$.} %
\label{figeqmajo}
\end{center}
\end{figure}

For some unknown reason, Majorana did not study further his model
of ferromagnetism, probably regarding it
incomplete,\footnote{Contrary to several other topics, none about
ferromagnetism was reported in the {\it Volumetti}
\cite{Volumetti}, this being an indication of the fact that the
author did not regard his calculations as conclusive (see the
Preface of Ref. \cite{Volumetti}).} but he nevertheless considered
some applications to ferromagnetic materials with different
geometries, corresponding to different numbers $i$ of aligned
spins on a total of $n$ and different numbers $a$ of nearest
neighbors. Here we only mention these peculiarities, leaving aside
less interesting numerical calculations of several quantities
appearing in the model \cite{Quaderni}.

\begin{figure}[t]
\begin{center}
\begin{tabular}{c}
\epsfysize=1.5truecm
\epsffile{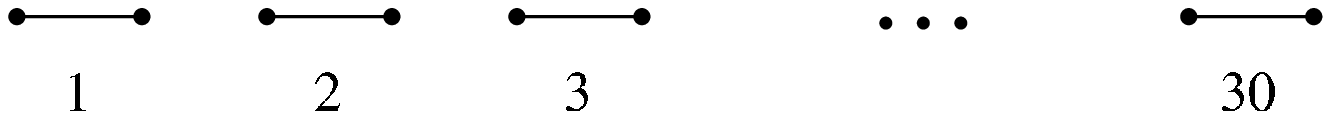} \\
\epsfysize=1.4truecm
\epsffile{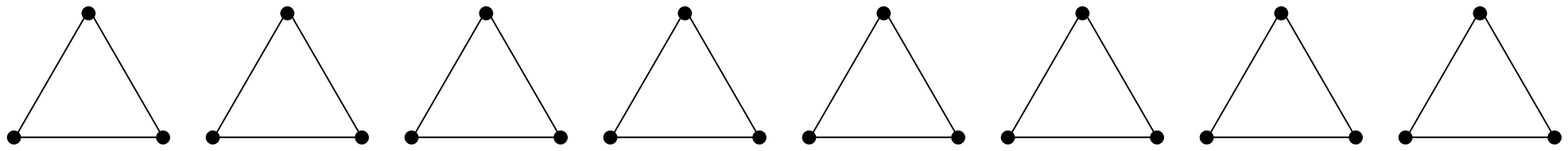} \\
\epsfysize=4truecm
\epsffile{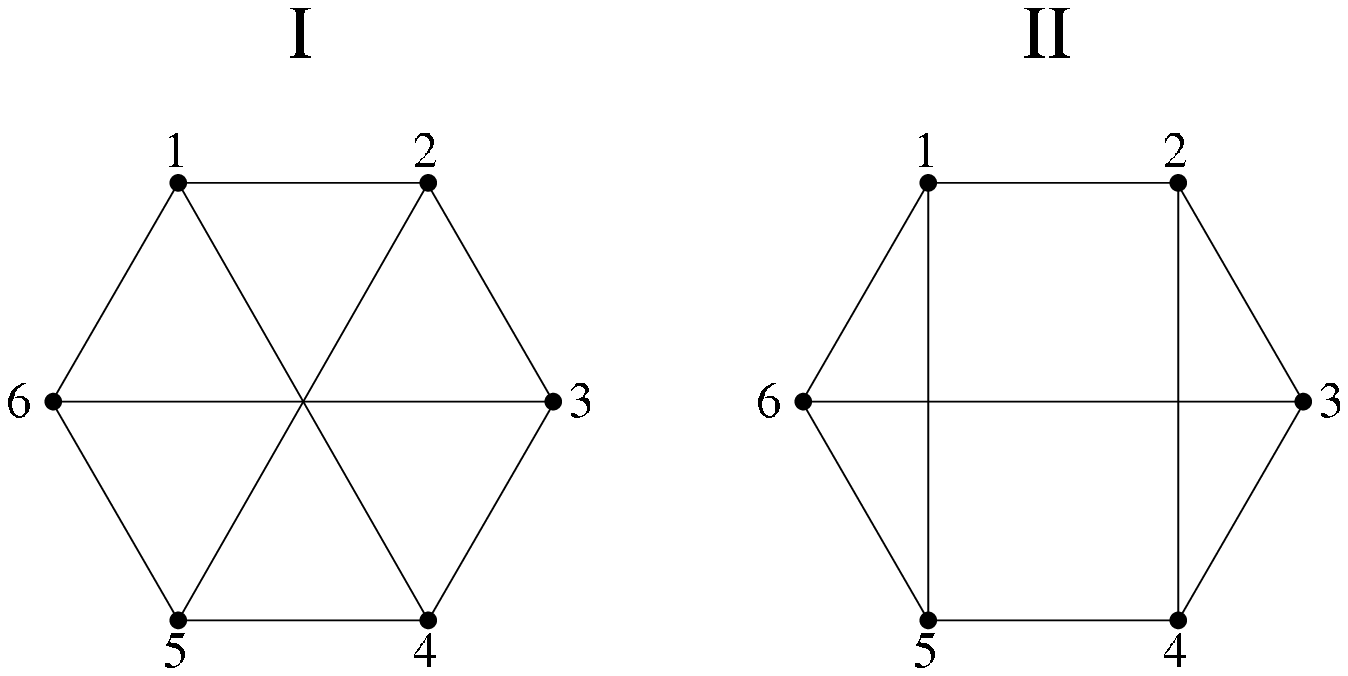} \\
\end{tabular}
\caption{Some applications, with different geometries, of the
Majorana model of ferromagnetism. From the top: (a) geometric
arrangement of $n=30$ elementary magnets (with $a=1$ nearest
neighbor) into a linear chain; (b) triangular chain of $n=24$
elementary magnets having $a=2$ nearest neighbors; (c) possible
geometric arrangements of $n=6$ elementary magnets, each of them
with $a=3$ nearest neighbors.} %
\label{figchains}
\end{center}
\end{figure}

The simplest example is that of a linear chain but, contrarily to
what usually considered (with $a=2$), Majorana used $a=1$ as
depicted in Fig. \ref{figchains}a. Explicit numerical calculations
for this case referred to $n=60$ and $i=10$.

The two dimensional case, with $a=2$, was instead implemented by
means of a triangular geometry, as showed in Fig.
\ref{figchains}b, with $n=24$ and $i=6$.

Finally Majorana also considered the case with $a=3$ (and $n=6$,
$i=3$), realized with two possible, different geometries as
illustrated in Fig. \ref{figchains}c.

While it is unquestionable that such examples are the simplest
ones in the Majorana model, it is unknown to what particular
materials Majorana thought when discussing them. Note, in fact,
that the only elements known at that time to exhibit ferromagnetic
properties were iron, cobalt and nickel, whose crystal structures
are apparently different from those considered by Majorana (body
centered cubic for Fe, face centered cubic for Fe and Ni,
hexagonal for Co). However, the realization of unusual geometric
arrangements, even in usual materials, could be well explained in
terms of different interactions responsible of the effective
equilibrium positions of the elementary magnets. In this respect
it is quite illuminating the case of a ``triatomic molecule
forming an equilateral triangle'' in a state analogous to that of
ferromagnetism, studied by Inglis \cite{Inglis3} and, especially,
that of pyrrhotite (an iron sulfide mineral with a variable iron
content). The same author modelled this element in terms of ``one
$d$-electron (or several independent $d$-electrons) per atom, with
fairly large spin-orbit interaction (but not as large as the
interatomic exchange energies), in a hexagonal crystal which has
the iron atoms in successive principal planes not directly
opposite one another, but staggered, or else far apart''
\cite{Inglis2}. If the reasoning of Inglis is really similar to
that of his friend Majorana, the model prepared by this scholar
was probably intended for providing a suitable framework for
explaining peculiar phenomena (such as ferromagnetic anisotropy
and others) not considered in the bare Heisenberg model. However,
since no further certain indications exist in the Majorana
notebooks, we prefer to not lucubrate on this, but rather stop
here our analysis.

\section{Conclusion}

\noi In this paper we have set forth and analyzed a theory of
ferromagnetism developed by Majorana as early as the beginnings of
1930s, but never published by him. Such a  theory differs
substantially (in the methods employed and in the results
obtained) from the well-known Heisenberg theory of 1928, although
the general framework is the same. In fact, in both cases, the
guidelines are those of the statistical Ising model of 1925, and
the forces responsible of ferromagnetism are just derived from the
quantum mechanical exchange interactions. Differently from
Heisenberg, however, Majorana described the system of
spin-up/spin-down elementary magnets in terms of wavefunctions
written in the form of a Slater determinant (as did by Bloch and
by Slater in 1929), in order to take directly into account the
constraints imposed by the Pauli exclusion principle.

The distribution function $y(N)$ for the configurations of the
system of spins modelling the ferromagnet is, then, explicitly
evaluated in the continuum limit (that is, for a continuous
distribution of the energy levels of the system, as in the
Heisenberg model). Indeed, from the observation that given
configurations of spin-up/spin-down elementary magnets are
statistically equivalent to those where a spin-up is replaced by a
spin-down (or viceversa) due to thermal fluctuations, Majorana
calculates the probability for those configurations of spins and,
by solving a differential equation, he obtains the expression for
$y(N)$, from which the partition function of the system may be
deduced. It is also interesting to mention that Majorana gives a
series expansion of the distribution function obtained which goes
well beyond the usual Gaussian approximation (with terms up to
order 10).

The introduction of an external magnetic field acting on the
elementary magnets allows the calculation of the appropriate
partition function of the system and, from a maximum problem (on
the probability for realizing a given magnetization), a non-linear
equation for the mean magnetization $S$ of the system is derived.
Although such equation is different from that deduced by
Heisenberg (and that deduced early by Weiss), it nevertheless
admits a non trivial solution $S \neq 0$ even in the absence of an
external magnetic field, provided that the temperature of the body
is lower than a critical (non-vanishing) value. The main features
of ferromagnetism are, thus, obtained in a more transparent manner
with respect to Heisenberg and subsequent (for instance, Bloch)
derivations.

In addition, and differently from these previous models, a
condition on the number of nearest neighbors {\it directly}
follows in the Majorana theory (instead, in the Heisenberg theory,
for example, a condition on this number is only supposed on the
basis of available experimental data), anticipating in some
respects a later key result by Peierls.

Finally, some applications of the theory to peculiar arrangements
of the elementary magnets (linear chain, triangular chain, etc.)
are as well considered by Majorana.

From what presently discussed, then, it comes out clear the
relevance of the contribution by Majorana to the theory of
ferromagnetism which, as happened for several other contributions
by this scholar in different fields of physics research
\cite{ADP}, unfortunately remained unknown until now. This
evidently urges to look into the several thousands of pages of his
personal research notes \cite{Volumetti} \cite{Quaderni}, whose
careful study has been undertaken only recently and which will
certainly deserve, once more, further interesting results.



\end{document}